\documentstyle[12pt]{article}
\begin{document}

\begin{flushright}
UT-Komaba 96-7
\end{flushright}


\begin{center} 
{\Large{\bf Statistics of Holons and Spinons in the t-J Model }}  
\vskip 0.3in
 {\Large  Ikuo Ichinose$^{\star}$,\footnote{e-mail address: 
ikuo@hep1.c.u-tokyo.ac.jp}
  Tetsuo Matsui$^{\dagger}$,\footnote{e-mail address: 
matsui@phys.kindai.ac.jp}
   and Kazuhiko Sakakibara$^{\ast}$\footnote{e-mail address: 
sakaki@center.nara-k.ac.jp}}\\
\vskip 0.2in
 $^{\star}$Institute of Physics, University of Tokyo, Komaba, Tokyo, 153 Japan  \\
 $^{\dagger}$Department of Physics, Kinki University, Higashi-Osaka,  577 Japan \\
$^{\ast}$Department of Physics, Nara National College of Technology, Yamatokohriyama,
639-11 Japan
\end{center}
\begin{center} 
\begin{bf}
Abstract
\end{bf}
\end{center}
We study the statistics of holons and spinons in the 2-dimensional t-J model
of high-Tc superconductors by applying the gauge theory of  
separation phenomena that we have developed recently.
This study is motivated by the observation that, near the half filling, the 
spin degrees of 
freedom of the t-J model are correctly described by  bosonic variables
as in the slave-fermion representation, while, at intermediate hole concentrations,
spinons  behave  as   fermions as in the slave boson representation, supporting  
a large Fermi surface.
In the previous papers, we   studied   
the  charge-spin separation (CSS) in high-Tc superconductors and
 the particle-flux  
separation (PFS) in the fractional quantum Hall effect (FQHE), and showed that both  
of them are understood as   
phase transitions   of certain  gauge theories into their deconfinement phases,
 where the gauge fields are nothing but the phases 
of mean fields sitting on links. For example,
the CSS occurs when the   dynamics of the gauge field that glues a holon and
 a spinon is relaized in the deconfinement phase; holons and spinons 
appear as quasiexcitations. 
In the present paper, we  start with the slave-boson representation
of the t-J model, in which an electron is a composite of a fermionic spinon 
and a bosonic holon. Then we represent each fermionic spinon  as a composite of 
a new boson and one (or odd number of)  Chern-Simons (CS) flux.
  A new  PFS, similar to the PFS in FQHE, occurs
when the dynamics of   
gauge field that glues each new bosonic spinon and  a  CS flux lies 
in the deconfinement phase.
 The following four  phases are possible; 
(1) electron phase (no separations), 
(2) anormalous metallic phase with fermionic spinons and bosonic holons (CSS), 
(3) double-boson phase (bosonic holons and spinons) with  antiferromagnetic long-range 
order (CSS and PFS), and  (4)  bosonic electron phase (PFS).
We discuss the interplay of CSS and PFS.
Especially, it is  concluded  that the spinons  behave   as  bosons   rather than 
fermions near the half filling due to the PFS.

\newpage 
\setcounter{footnote}{0}
\section{Introduction}

Since the discovery of high-Tc superconductivity,  strongly-correlated
electron systems are one of the most interesting topics in condensed 
matter physics.
Especially, the metallic phase of high-Tc superconductors has various 
anormalous properties.
The anormalous experimental observations can be consistently explained by assuming
that charge and spin degrees of freedom of electrons (holons and spinons, 
respectively) move independently, i.e.,
the charge-spin separation (CSS) takes place \cite{css}.
In the mean-field (MF) theory of the t-J model in the slave-boson
or slave-fermion representation \cite{meanfield1,meanfield2}, the CSS appears quite naturally.
However, the MF calculations themselves are not sufficient, because the phase
degrees of freedom of the MF's sitting on links behave like gauge fields
that glue holons and spinons,
and careful study of the dynamics of these effective gauge fields is
required to justify the results of MF theory.

In the previous papers \cite{IMCSS}, two of the present authors 
examined the phase structure of the above effective gauge fields.
As we showed there, the CSS can be understood  as a 
deconfinement phenomenon, i.e., gauge dynamics operating between holons and spinons 
is so weak that holons and spinons are deconfined (liberated).  
There exists a confinement-deconfinement
 (CD) phase transition at certain  critical temperature, $T_{\rm CD}$, and
the CSS occurs at {\em low} temepratures ($T$) below $T_{\rm CD}$ (deconfinement
 phase),
while at $T > T_{\rm CD}$, holons and spinons are confined and appear
solely as electrons (confinement phase).

In the present paper, we shall address yet another interesting problem 
of the strongly-correlated
electron systems, i.e., statistics transmutation of the quasiexcitations in the CSS phase.
Actually, in high-Tc superconductors, it is generally expected that, at intermediate 
hole concentrations,  spinons behave 
as fermions, while, near the half filling, spin dynamics 
is well described by bosonic variables as in the $O(3)$ nonlinear $\sigma$-model.
In two-dimensional systems, a statistics transmutation of particles is properly described
by the Chern-Simons (CS) gauge theory.
For example, let us recall the fractional quantum Hall effect (FQHE)  at 
filling fractions $\nu = 1/(2n+1)$,
 which is another interesting two-dimensional strongly-correlated electron system.
The FQHE is quite successfully described by the CS Ginzburg-Landau theory \cite{Girvin},
which takes a form of the CS gauge theory coupled with a bosonized electron field.
A FQH state is characterized here as a Bose condensation of these bosonized electrons.

Another example is quasiexcitations in the half-filled Landau level. 
Jain \cite{Jain} proposed
the idea of composite fermions (CF); an electron is a composite of a CF and
two units of solenoidal CS fluxes. By assuming that CF's and fluxes move independently,
one may develop a MF theory, in which fluxes are globally cancelled by
the external magnetic field, and the resulting system is a collection of
quasifree CF's weakly interacting with a  gauge field.  The perturbative analyses of such a system
give rise to interesting results \cite{NW}. 
In a previous paper \cite{IMPFS}, we called this possible separation phenomena,  ``particle-flux
separation" (PFS), due to its close resemblance to the CSS, and
 studied the mechanism of PFS by applying  the gauge-theoretical method similar to the case of CSS.   We found that there is a CD phase transition at some 
critical temperature $T_{\rm PFS}$;
At $T < T_{\rm PFS}$, the system is in the deconfinement phase and  electrons
are dessociated into CF's and fluxes, i.e., the PFS
takes place, while at $T >  T_{\rm PFS}$, the system is in the confinement phase
and CF's and fluxes are confined into electrons.

Being motivated by these analyses,   we shall study statistics transmutation of the
quasiexcitations  of the t-J model by employing the 
CS gauge-field formalism together with  the above gauge theory of sepatation phenomena.  

In Sect.2, we start with the slave-boson representation of the t-J model,
in which an electron $C_{x\sigma} ( x$; site, $\sigma$; spin) 
is expressed as a composite of a fermionic spinon $f_{x \sigma}$
and a bosonic holon $b_x$. As in the usual MF theory, we decouple the model
by intorducing auxiliary MF's sitting on links. Since their phases are gauge fields
that glue holons and spinons, analysis of the phase dynamics is crucial
to study CSS.   
By using the CS gauge theory, we furthermore express a fermionic 
spinon  as a composite of a new boson (bosonic spinon) $\phi_{x \sigma}$
 and  odd number of CS flux quanta $W_x$
to rewrite the Hamiltonian in the double-boson representation.
Next,  we   introduce  another auxiliary
link field, the phase of which is a new gauge field that glues a  bosonic spinon 
and a CS flux.
We study its dynamics, under the effects of the other gauge 
fields introduced to study CSS, 
following our general method to the separation phenomena.
Phase structure of this gauge dynamics  is closely related to the statistics of 
spinons. 
If this gluing gauge field is in a deconfinement phase, bosonic spinons
and CS fluxes are deconfined and move independently; spinons behave  as 
bosons rather than  fermions; statistics of the spinon is transmuted.
On the other hand, if it is in a confinement phase, composites do not break;
quasiexcitations are original fermionic spinons themselves in the slave-boson 
formalism from which we have started.
In the rest of the present paper, we shall call the above separation of a 
bosonic spinon  and a CS flux in the t-J model
also  PFS due to its common nature to the PFS at FQHE.
Schematically, one may summerize the relation as 
\begin{eqnarray}
C_{x\sigma} &=& b^{\dagger}_{x} \times  f_{x \sigma} 
\rightarrow b^{\dagger}_{x} +  f_{x \sigma}, \ ({\rm CSS})
\nonumber\\
f_{x\sigma} &=& \phi_{x \sigma} \times W_x^{\dagger} 
\rightarrow \phi_{x \sigma} +  W_x^{\dagger},  \ ({\rm PFS})
\end{eqnarray}
where the symbol $ ``\times"$ means confinement and $``+"$ means deconfinement
(separation).

In Sect.2.2, we list up the possible four phases which the system may exhibit, using
a general argument based on the gauge symmetries of the double-boson t-J model.
According to whether the CSS and/or the PFS take place, one expects the following
four phases;  \\\noindent
(1) the electron phase (no separation),\\\noindent 
(2) the anormalous metallic phase with fermionic spinons and bosonic holons
(CSS),\\\noindent 
(3) the double-boson phase (bosonic holons and spinons) with  antiferromagnetic long-range 
order (CSS and PFS),\\\noindent 
(4) the bosonic electron phase (PFS). 

In Sect.3.1, we formulate a MF theory by integrating out bosonic holons and spinons using the hopping expansion. This determines the magnitudes of link 
fields; they develop nonvanishing values at sifficiently low $T$. 
Thus the phases of link variables can be introduced. They are
gauge fields and  play an important role  at low energies. 
In Sect.3.2, we derive an effective gauge theory of the above gauge fields 
by the hopping expansion. Its action has a form similar to the well-known 
lattice gauge theory developed for strong interactions.

In Sect.4, the phase structure of the effective gauge theory is studied explicitly.
In Sect.4.1, we first investigate a canonical gauge model from a general
point of view.  According to the   Polyakov-Susskind
theory of CD transition, we  map  the gauge model to the (an)isotropic
classical XY spin model. 
In Sect.4.2 we  apply the results obtained there to the effective gauge 
theory of Sect.3.2.
We show that, both for CSS and PFS, there exist    CD phase transitions 
and derive  the equations that 
determine the critical temperatures $T_{\rm CSS}$, $T_{\rm PFS}$; 
the CSS and PFS take  place at  $T < T_{\rm CSS}$ and $T < T_{\rm PFS}$, respectively.
Furthermore, we give a plausible argument that the spinons behave  as bosons
rather than fermions {\it near the half filling}, in accordance with
the  slave-fermion representation.

In Sect.5, we  present conclusions and point out the remaining problems. 
Generally, one may conceive the following two representations of 
fermionic electron operator;
\begin{eqnarray}
&& {\rm Double \  Boson:}\nonumber\\
&&  C_{x\sigma} = b^{\dagger}_{x} \times  f_{x \sigma} =
b^{\dagger}_{x} \times  \phi_{x \sigma} \times W_x^{\dagger},
\label{db}\\
&& {\rm Double \ Fermion:}\nonumber\\
&&  C_{x\sigma} = \psi^{\dagger}_{x} \times  a_{x \sigma} =
\psi^{\dagger}_{x} \times  \eta_{x \sigma} \times W_x^{\dagger},
\label{df}
\end{eqnarray}
where $\psi_x$ is the fermionic holon operator and $a_{x\sigma}$
is the bosonic spinon operator in the slave-fermion representation and 
$\eta_{x \sigma}$ is the fermionic spinon operator.
In the present paper, we mainly focus on the statistics transmutation of 
spinons. As explained above,   the bosonic spinons 
are expected to be superior assignment 
than the fermionic spinons at small hole concentrations. We
start with the double boson representation (\ref{db}) rather than
the double fermion representation (\ref{df}), and demonstrate this superiority 
via the PFS.
It is obvious that the statistics of holons is also an important problem, and one should
study this problem in order to obtain a coherent picture of quasiexcitations in the t-J model.
For example, if  holons behave as fermions at low $T$ and intermediate
hole concentrations, a hole pair with electric charge $+2e$, 
{\em not a single hole}, is a natural candidate for the condensed
objects that develop the superconductivity \cite{imsuper}. 
All these problems are under study, and results will be reported in  future publications.

\setcounter{equation}{0}
\section{The t-J model in double-boson representation }

In Sect.2.1, we rewrite the t-J model in several steps to
arrive at the double boson representation, (\ref{H2}) below, 
which is useful to discuss the CSS and PFS.
In Sect.2.2, we present a general discussion on what kinds of phases
this model may realize, being based on the two local $U(1)$ gauge symmetries 
((\ref{symmetry1}) and (\ref{symmetry2})) of the model. 

\subsection{Double-boson representation}
The t-J model on a two-dimensional lattice is defined by the following Hamiltonian and the local constraint,
\begin{eqnarray}
H_{tJ}&=&-t\sum_{x,i,\sigma}C^{\dagger}_{x+i,\sigma}C_{x \sigma}
+J\sum_{x,i}\left({\mbox{\boldmath $S$}}_{x+i}\cdot {\mbox{\boldmath $S$}}_{x}
- \frac{1}{4} n_{x+i} n_{x} \right), \nonumber  \\
n_x & \equiv & \sum_{\sigma}C^{\dagger}_{x \sigma} C_{x \sigma} \leq 1,
\label{HtJ}
\end{eqnarray}
where $C_{x \sigma}$ is the electron annihilation operator at site $x$ and spin 
$\sigma (= 1,2) $,
and the spin operator is given by
{\boldmath $S$}$_x= C^{\dagger}_{x}${\boldmath $\sigma$}$C_x /2$ 
with the Pauli matrices {\boldmath $\sigma$}.
The suffix $i (= 1,2)$ denotes the lattice directions and also the unit lattice 
vectors.
In the slave-boson representation the electron operator is written as
\begin{equation}
C_{x \sigma}=b^{\dagger}_xf_{x \sigma}, 
\label{slaveb}
\end{equation}
where $b_x$ is the bosonic holon
operator and $f_{x \sigma}$ is the fermionic spinon operator. 
 The local constraint
in (\ref{HtJ}) is expressed as
\begin{equation}
(f^{\dagger}_{x 1}f_{x 1} + f^{\dagger}_{x 2}f_{x 2} + b^{\dagger}_x b_x-1)
|\mbox{phys}\rangle=0,  
\label{local}
\end{equation}
which excludes  double occupancy of electrons at each $x$.
By substituting (\ref{slaveb}) into (\ref{HtJ}), 
the Hamiltonian is rewritten as
\begin{eqnarray}
H&=&-t\sum (b^{\dagger}_{x+i}f^{\dagger}_{x \sigma}f_{x+i,\sigma}b_x+\mbox{H.c.})
-{J\over 2} (\sum  f^{\dagger}_{x \sigma}\tilde{f}_{x+i,\sigma}) 
 (\sum \tilde{f}^{\dagger}_{x+i,\sigma'}f_{x \sigma'}) \nonumber   \\
& -&\mu_f\sum f^{\dagger}_{x \sigma}f_{x \sigma}  
 -\mu_b\sum b^{\dagger}_xb_x- \sum \lambda_x\Big(\sum f^{\dagger}_{x \sigma}
 f_{x \sigma}+b_x^{\dagger}b_x-1\Big), \nonumber   \\
\tilde{f}_{x \sigma} &\equiv &
\sum_{\sigma'}\epsilon_{\sigma \sigma'}f^{\dagger}_{x \sigma'}
\label{Hsb},
\end{eqnarray}
where  
$\epsilon_{\sigma \sigma'}$ is the antisymmetric tensor.  We added the chemical 
potential terms and the Lagrange multiplier $\lambda_x$ for the local constraint 
for later convenience. $\mu_{b}$ and $\mu_{f}$ are chosen so that the hole concentration is 
$\delta$, i.e., $\langle b^{\dagger}_x b_x\rangle = \delta$, and so $  \langle\sum_{\sigma} 
f^{\dagger}_{x \sigma} f_{x \sigma}\rangle =\rho=1 - \delta$.

By using the Hubbard-Stratonovich transformation in path-integral formalism,
let us rewrite (\ref{Hsb}) in the  form used in a MF theory \cite{meanfield1}.
Explicitly, by introducing auxiliary complex link variables, $\chi_{x i}$
and $D_{x i}$, that are {\it fluctuating} ``MF's", we get \cite{decoupling}
\begin{eqnarray}
H&=&\sum \Big[ {3J\over 8}|\chi_{x i}|^2+{1\over 2J}|D_{x i}|^2
\Big]-\sum\Big[\chi_{x i}\Big({3J\over 8}\sum f^{\dagger}_{x+i,\sigma}
f_{x \sigma} 
+t b^{\dagger}_{x+i}b_x\Big) +\mbox{H.c.}\Big]  \nonumber \\
&-&{1\over 2}\sum 
[ D_{x i} f^{\dagger}_{x \sigma}\tilde{f}_{x+i,\sigma} +\mbox{H.c.} ]
+ {8t^2 \over 3J}\sum b^{\dagger}_{x+i}b_{x+i}b^{\dagger}_xb_x  \nonumber  \\
&-& {3J\over 8}\sum f^{\dagger}_{x+i,\sigma}f_{x+i,\sigma'}
f^{\dagger}_{x \sigma'}f_{x \sigma}
-\mu_f \sum f^{\dagger}_{x \sigma}f_{x \sigma}
-\mu_b\sum b^{\dagger}_xb_x  \nonumber  \\
&-&  \sum \lambda_x\Big(\sum f^{\dagger}_{x \sigma}
 f_{x \sigma}+ b_x^{\dagger}b_x-1\Big).
\label{Hdec1}
\end{eqnarray}
By differentiating $H$, we get the relations, 
\begin{eqnarray}
\frac{3J}{8} \langle\bar{\chi}_{x i}\rangle &=&  
\langle \frac{3J}{8} \sum_{\sigma}
f^{\dagger}_{x+i,\sigma}
f_{x \sigma} +t b^{\dagger}_{x+i}b_x \rangle, \nonumber\\
\frac{1}{2J} \langle\bar{D}_{x i}\rangle &=& \frac{1}{2} \langle \sum_{\sigma} f^{\dagger}_{x\sigma}\tilde{f}_{x+i,\sigma}\rangle.
\end{eqnarray}
So $\chi_{x i} $ is the hopping amplitude
of holons and spinons, while  $D_{xi}$ is the amplitude of resonating valence bonds of antiferromagnetism.
In the usual MF approximation, one  simply
determines the values of MF's,
$\langle\chi_{x i}\rangle$ and $\langle D_{x i}\rangle$ as functions of $\delta$ and  $T$,
assuming some spatial periodicity. The effects of fluctuations around
 these MF's, particularly the CD transition and the CSS, are studied in  Ref.
 \cite{IMCSS} based on this Hamiltonian \cite{note1}. 

According to the lattice CS gauge theory \cite{fradkin}, let us introduce a bosonic spinon field $\phi_{x \sigma}$ \cite{sbsftrans}, 
\begin{eqnarray}
\phi_{x \sigma}&=&e^{-iq\sum_{y} \theta (x-y) (\hat{\rho}(y) - \rho)}
f_{x \sigma}, \nonumber   \\
\hat{\rho}(x)&=&\sum_{\sigma}f^{\dagger}_{x \sigma}f_{x \sigma}
=\sum_{\sigma}\phi^{\dagger}_{x \sigma}\phi_{x \sigma}, 
\label{phi}
\end{eqnarray}
where $q$ is an odd integer and $\theta (x)$ is the multi-valued 
 lattice angle function with $\theta(0)=0$.
One can  verify that $\phi_{x \sigma}$, $\phi_{y \sigma'}$ and their  Hermitian
conjugate operators satisfy
the bosonic commutation relations for $x\neq y$ and the fermionic anticommutation 
relations  for $x=y$. These relations imply that  $\phi_x$ describes
 hard-core bosons.
By substituting (\ref{phi}), $H $ becomes   
\begin{eqnarray}
H &=&\sum \Big[ {3J\over 8}|\chi_{x i}|^2+{1\over 2J}|D_{x i}|^2
\Big]-\sum\Big[\chi_{x i}\Big({3J\over 8}\sum \phi^{\dagger}_{x+i,\sigma}
e^{-iq\sum_y \nabla_i \theta (x-y)(\hat{\rho} (y)-\rho)}
 \phi_{x \sigma} \nonumber \\
&+&tb^{\dagger}_{x+i}b_x\Big)+\mbox{H.c.}\Big]  
-{1\over 2}\sum 
[D_{x i} \;\phi^{\dagger}_{x \sigma}\tilde{\phi}_{x+i,\sigma}+\mbox{H.c.}]
+ {8t^2 \over 3J}\sum b^{\dagger}_{x+i}b_{x+i}b^{\dagger}_xb_x  \nonumber  \\
&-&{3J\over 8}\sum \phi^{\dagger}_{x+i,\sigma}\phi_{x+i,\sigma'}
\phi^{\dagger}_{x \sigma'}\phi_{x \sigma}
-\mu_\phi\sum \phi^{\dagger}_{x \sigma}\phi_{x \sigma}
-\mu_b\sum b^{\dagger}_xb_x  \nonumber  \\
&-& \sum \lambda_x\Big(\sum \phi^{\dagger}_{x \sigma}
 \phi_{x \sigma}+b_x^{\dagger}b_x-1\Big),
 \label{Hdec12}
 \end{eqnarray}
where $\nabla_i$ is the difference operator on the lattice, and
we put $\mu_{\phi} = \mu_f$.

From the $\phi_{x \sigma}$-hopping term in (\ref{Hdec12}), it is obvious that
the bosonic spinons move  in the combined field of  $\chi_{xi}$ 
and the CS magnetic field $B^{CS}_x$,
\begin{eqnarray}
B^{CS}_x&\equiv&\epsilon_{ij}\nabla_iA^{CS}_{x j}=2\pi q\hat{\rho}(x) \nonumber   \\
A^{CS}_{x i}&=&q\sum_y \nabla_i\theta(x-y)\hat{\rho}(y),
\label{CSB}
\end{eqnarray}
where we have used the fact 
\begin{equation}
\nabla_i\theta (x)=2\pi\epsilon_{ij}\nabla_jG(x),
\label{Green}
\end{equation}
and $G(x)$ is the two-dimensional lattice Green function.

As mentioned in the Introduction, two of the present authors discussed quasiexcitations 
in the half-filled Landau level by using the gauge theory similar to that for  CSS.
There we introduced a gauge field which glues a CF and CS fluxes
and studied its dynamics. We found that the PFS occurs  at $T < T_{\rm PFS}$, 
where these CF's are stable quasiexcitations.

Here, we apply the same method and introduce  auxiliary complex variables
 $V_{xi}$ on links by the Hubbard-Stratonovich transformation. Then $H$
is rewritten as follows;
\begin{eqnarray}
H &=&\sum \Big[ {3J\over 8}|\chi_{x i}|^2+{1\over 2J}|D_{x i}|^2 
+|V_{x i}|^2 \Big]-\Big[\sum\chi_{x i}\Big({3J\over 8\gamma}
\sum \phi^{\dagger}_{x+i,\sigma}
V_{x i} \phi_{x \sigma} \nonumber \\
&+& t b^{\dagger}_{x+i}b_x\Big)   
+{1\over 2}\sum 
 D_{xi}\;\phi^{\dagger}_{x \sigma}\tilde{\phi}_{x+i,\sigma}+ 
\gamma\sum V_{x i}W^{\dagger}_{x+i}W_x+\mbox{H.c.} \Big] \nonumber   \\
&-&  \sum  g^2_{\phi} \phi^{\dagger}_{x+i,\sigma}\phi_{x+i,\sigma'}
\phi^{\dagger}_{x \sigma'}\phi_{x \sigma}
+{8t^2 \over 3J}\sum b^{\dagger}_{x+i}b_{x+i}b^{\dagger}_xb_x   
-\mu_\phi\sum \phi^{\dagger}_{x \sigma}\phi_{x \sigma}  \nonumber  \\
&-& \mu_b\sum b^{\dagger}_xb_x  
- \sum \lambda_x\Big(\sum \phi^{\dagger}_{x \sigma}
 \phi_{x \sigma}+b_x^{\dagger}b_x-1\Big),
 \label{H2}
 \end{eqnarray}
where
\begin{eqnarray}
W_x&\equiv&\exp [-iq\sum_y \theta(x-y)(\phi^{\dagger}_{y \sigma}\phi_{y \sigma} -\rho) +i\omega_x],
\label{W} \\
g^2_{\phi}&\equiv&{3J\over 8}+\Big({3J\over 8\gamma}|\chi_{x i}|\Big)^2, 
\label{gphi} 
\end{eqnarray}
and we have introduced the parameter $\gamma$ for dimensional reason.
In (\ref{W}), $\omega_x$ is an arbitrary function, which corresponds to the longitudinal 
part of the ``CS gauge field" (see Ref.\cite{IMPFS}).

The partition function $Z \equiv {\rm Tr}\exp(-\beta H)$ of the system is expressed 
in the path-integral formalism as 
\begin{eqnarray}
Z &=&  \int [ d\phi ][ db ][ d\chi ][ dD ][ dV][
d\omega] \exp A, \nonumber\\
A & = & \int^{\beta}_0 d\tau \;\Big[-\sum\phi^{\dagger}_{x \sigma}
\partial_{\tau}\phi_{x \sigma}-\sum b^{\dagger}_x \partial_{\tau}b_x-H\Big]  
\label{Adef}
\end{eqnarray}
where the imaginary time $\tau$ runs from $0$ to $\beta \equiv 1/T$.
Here $[db][d\chi][dD][dV]$ are usual path integrals of complex variables, and
$[d\omega]$ is the path integral for $0 < \omega_x(\tau) < 2\pi$ which respects
the gauge symmetry [$U(1)_{\rm PFS} $ explained below].
The integral $[d\phi]$ is
ambiguous since no simple coherent states of hard-core bosons are known. We
take a possible option that $\phi_{x}(\tau)$ is a complex number and
the measure includes a factor $\lim_{\lambda_\phi \rightarrow \infty}
 \exp [-\int d\tau \lambda_\phi \sum f(\bar{\phi}_x\phi_x)]$, where 
$f(x)$ is some smooth function which is zero for $x =0,1$ and positive otherwise.
This option is too formal generally, but is
sufficient for the hopping expansion that we shall use;  We can calculate
each terms of hopping expansions systematically respecting the hard-core nature
(See Sect.3 for details).


\subsection{Gauge symmetries $U(1)_{\rm CSS}$,  $U(1)_{\rm PFS}$
and possible phases }

The Hamiltonian (\ref{H2}) is invariant under the following two kinds of
time-independent local U(1) gauge transformations;
\begin{eqnarray}
U(1)_{\rm CSS};\ \ \ &&(\phi_{x \sigma},b_x) \rightarrow  
e^{i\alpha_x}(\phi_{x \sigma},b_x),  \nonumber   \\
&&\chi_{x i} \rightarrow  
e^{i\alpha_{x+i}}\chi_{x i}e^{-i\alpha_x}  \nonumber   \\
&&D_{x i} \rightarrow   e^{i\alpha_{x+i}}D_{x i}e^{i\alpha_x},
\label{symmetry1}
\end{eqnarray}
and 
\begin{eqnarray}
 && \nonumber\\
U(1)_{\rm PFS};\ \ \ &&(\phi_{x \sigma}, W_x) \rightarrow   e^{i\beta_x}
 (\phi_{x \sigma},W_x)  \nonumber  \\
&& \omega_x \rightarrow \omega_x + \beta_x  \nonumber  \\
&&V_{x i} \rightarrow e^{i\beta_{x+i}}V_{x i}e^{-i\beta_x}.
\label{symmetry2}
\end{eqnarray}

We shall derive an effective gauge theory of (\ref{H2}) in Sect.3 and discuss its
dynamics in Sect.4. Since the basic characteristics  of general gauge dynamics are 
determined sorely from the knowledge of gauge symmetry,
it will be helpful to discuss the possible phases of the t-J model, based on
the gauge symmetries $U(1)_{\rm CSS}$ and $U(1)_{\rm PFS}$, before going into details.

Generally speaking, dynamics of gauge theory is categorized into two phases; confinement phase and 
deconfinement phase.
In the confinement phase, fluctuations of   gauge field are very large, i.e., it is a
disordered phase of the gauge symmetry, and only charge-neutral bound objects
can appear as physical excitations.  Contents of quasiparticles are quite different from the original ``elementary"
charged particles, which appear in the Hamiltonian.
On the other hand, in the deconfinement phase,  fluctuations of   gauge
field is not so strong, and charged particles appear as physical
excitations.

Since our system possesses  two independent gauge symmetries, 
$U(1)_{\rm CSS}$ and $U(1)_{\rm PFS}$, there are  totally  four possible 
phases.
They are   listed up with  their physical  picture as follows; 
\begin{enumerate}
\item ($U(1)_{\rm CSS}$, $U(1)_{\rm PFS}$)=(confinement, confinement) \\
Only charge-neutral bound states of $\phi_{x \sigma}$ and $b_x$ as well as
$\phi_{x \sigma}$  and $W_x$ are 
physical quasiexcitations. They are composites of $\phi_{x \sigma}$, $b_x$  and $W_x$,
or equivalently $f_{x \sigma} = W_x\phi_{x \sigma}$ and $b_x$,
which  are nothing but the electrons $C_{x\sigma} = b^{\dagger}_x f_{x \sigma}$.
Therefore, this phase is the electron phase.
In the t-J model,  it is expected to appear  at sufficiently high $T$.  
\item ($U(1)_{\rm CSS}$, $U(1)_{\rm PFS}$)=(deconfinement, deconfinement) \\
In this phase, bosonic excitations $\phi_{x,\sigma}$ and $b_x$ appear
as quasiexcitations.
Since fluctuations of link fields $\chi_{xi},D_{xi},V_{xi}$ are small,
a simple MF theory is applicable starting from the Hamiltonian (\ref{H2}).
It is expected that one obtains the results similar as in the MF theory
of  slave-fermion formalism of the t-J model for spinons in which 
bose statistics is assigned to spinons. 
The MF's $\chi_{xi}$ etc. may not have a simple uniform configuration
but $\sqrt{2}a$ or $2a$ periodicity as in the spiral state in the slave-fermion
MF theory, where $a$ is the lattice spacing.
The bose condensation of spinons $\phi_{x \sigma}$ at sufficiently low $T$
 corresponds to a long-range
order of antiferromagnetism.
Both the spiral state and the state with antiferromagnetic long-range order belong to this phase.
\item ($U(1)_{\rm CSS}$, $U(1)_{\rm PFS}$)=(deconfienement, confinement)  \\
$\phi_{x\sigma}$ and $b_x$ are not bound and so the CSS takes place, 
while $\phi_{x\sigma}$ and $W_x$ are bound
to form $f_{x\sigma}$.
Thus the quasiexcitations are fermionic spinons $f_{x\sigma}$ and bosonic holons $b_x$.
The usual slave-boson representation is useful for this phase, and there
appears a large Fermi surface consistent with the Luttinger theorem.
This phase corresponds to the anormalous metallic state in high-Tc superconductors.
\item ($U(1)_{\rm CSS}$, $U(1)_{\rm PFS}$)=(confienement, deconfinement)  \\
$\phi_{x\sigma}$ and $b_x$ are bound, 
while $\phi_{x\sigma}$ and $W_x$ are not bound. 
So the quasiexcitations are ``bosonic electrons", $b^{\dagger}_x\phi_{x,\sigma}$; 
which are bosons, each  having  spin $1/2$ and charge $-e$.
In  high-Tc superconductors,  no  states  corresponding to this phase seem to be observed.
However, if  holons cannot move freely by some effects like  localization, only relevent
 quasiexcitations are the composite spin excitations  localized at site $x$
and described by the operator 
{\boldmath $n$}$_x={1\over 2}\phi^{\dagger}_{x}${\boldmath $\sigma$}$\phi_x$.
 The spinons $\phi_{x\sigma}$ themselves cannot condense because $U(1)_{\rm CSS}$
 is in the confinement phase.
 Therefore, this phase corresponds to the insulator of cuprates with no long-range
 magnetic order. 
The dynamics of these massive spin excitations is well-described by the $CP^1$ or
 $O(3)$ nonlinear-$\sigma$ model (see detailed discussion in Ref.\cite{YTIM}).
 \end{enumerate}

In the rest of the present paper, we shall study which phase of these four will be
realized at each values  of $T$ and $\delta$.

\setcounter{equation}{0}
\section{Hopping expansion and effective gauge theory}

\subsection{Amplitudes of link fields}
We shall first study the amplitudes of link fields.
From the Hamiltonian (\ref{H2}), one may obtain a MF-like
Hamiltonian.
However, from the experience of MF studies of the t-J model,
it is expected that nonuniform configurations of the phases of link fields
may appear \cite{ps}.
This problem will be studied in some detail in the following subsection.
In this section we shall  focus on the single-link potential, 
which essentially
determines the amplitudes of  link fields.
For this purpose  we simple set in  (\ref{H2}),
 \begin{equation}
 \chi_{x i}=\chi_0, \; D_{x i}=D_0,  \; V_{x i}=V_0,
 \label{mean}
 \end{equation}
 where $ \chi_0, \; D_0$, and $V_0$ are assumed to be real constants
(this assumption is {\em not} essential in the following calculation). 
The MF Hamiltonian is given by
\begin{eqnarray}
H_{\rm MF} &=&\sum \Big[ {3J\over 8}|\chi_0|^2+{1\over 2J}|D_0|^2 
+|V_0|^2 \Big]-\Big[\sum\chi_{0}\Big({3J\over 8\gamma}
\sum \phi^{\dagger}_{x+i,\sigma}
 V_{0}\phi_{x \sigma} \nonumber \\
&+& t b^{\dagger}_{x+i}b_x\Big)   
+{1\over 2}\sum 
 D_{0}\;\phi^{\dagger}_{x \sigma}\tilde{\phi}_{x+i,\sigma}+ 
\gamma\sum V_{0}W^{\dagger}_{x+i}W_x+\mbox{H.c.} \Big] \nonumber   \\
&-&  g^2_{\phi} \sum \phi^{\dagger}_{x+i,\sigma}\phi_{x+i,\sigma'}
\phi^{\dagger}_{x \sigma'}\phi_{x \sigma}
+{8t^2 \over 3J}\sum b^{\dagger}_{x+i}b_{x+i}b^{\dagger}_xb_x   
-\mu_\phi\sum \phi^{\dagger}_{x \sigma}\phi_{x \sigma}  \nonumber  \\
&-& \mu_b\sum b^{\dagger}_xb_x  
- \sum \lambda_x\Big(\sum \phi^{\dagger}_{x \sigma}
 \phi_{x \sigma}+b_x^{\dagger}b_x-1\Big).
 \label{HMF}
 \end{eqnarray}
The propagators of the fields $\phi_{x \sigma}$ and $b_x$ in the hopping expansion
are obtained as follows;
\begin{eqnarray}
&&\langle b_x(\tau_1)b^{\dagger}_y(\tau_2) \rangle =\delta_{x y} G_b(\tau_1-\tau_2), 
\nonumber   \\
&&G_b(\tau)={e^{\mu_b\tau} \over 1-e^{\beta\mu_b}}  \; [\theta(\tau)+
e^{\beta\mu_b}\theta(-\tau)],  \nonumber  \\
&&\delta={e^{\beta\mu_b} \over 1-e^{\beta\mu_b}}, \nonumber   \\
&&\langle \phi_{x \sigma}(\tau_1)\phi^{\dagger}_{y \sigma'}(\tau_2)\rangle
=\delta_{\sigma \sigma'}\delta_{x y}G_{\phi}(\tau_1-\tau_2),  \nonumber   \\
&&G_{\phi}(\tau)={e^{\mu_{\phi}\tau} \over 1+e^{\beta\mu_{\phi}}} \;
[\theta(\tau)-e^{\beta\mu_{\phi}}\theta(-\tau)],  \nonumber  \\
&&1-\delta=2\cdot {e^{\beta\mu_{\phi}} \over 1+e^{\beta\mu_{\phi}}},
\label{propagator}
\end{eqnarray}
where we have treated $\phi_{x \sigma}$
as {\it hard-core} bosons as mentioned.

Here we  briefly comment on the treatment of the local-constraint Lagrange
multiplier $\lambda_x$ in $H_{\rm MF}$ (\ref{HMF}).
If the CSS takes place, the spinons and holons appear as quasiexcitations.
A straightforward loop calculation shows that,  due to
 the contribution from the loop diagrams of spinons and holons, fluctuations of $\lambda_x$ become massive, and so 
the local constraint becomes irrelevant at low energies \cite{IMCSS}.
This is consistent with the CSS.
On the other hand, if the CSS does not occur, the t-J model must be studied
in term of the original electrons, and the local constraint must be treated
faithfully.
One well-known example of such  treatments is the high-$T$ expansion.
As we are interested in the state of CSS in the present paper, we shall
ignore the local constraint in most of the later discussion.

Let us consider the effective potential $P$ of the link fields.
From (\ref{HMF}), on the tree level, we have 
\begin{equation}
P_{tree}={3J\over 8}\chi_0^2+{1\over 2J}D_0^2+V_0^2.
\label{Ptree}
\end{equation}
The $\phi_x$-hopping terms in (\ref{HMF}) give the following terms of $\chi_0$
and $D_0$ in the second-order of the hopping expansion;
\begin{equation}
P^{(2)}_{\phi}=-\frac{1}{2}\Big({3J\over 8\gamma}\Big)^2(1-\delta^2)\beta\chi_0^2V_0^2
-L(\delta)\beta D_0^2,
\label{P2phi}
\end{equation}
where $L(\delta)={\delta \over 4}\Big( \ln {1+\delta \over 1-\delta} \Big)^{-1}$,
and we have used the propagator (\ref{propagator}).  \\
Similarly, the $b_x$-hopping term gives rise to
\begin{equation}
P^{(2)}_b=-t^2\delta(1+\delta)\beta\chi_0^2.
\label{P2b}
\end{equation}

Contribution from the flux-hopping term $V_0W^{\dagger}_{x+i}W_x$ is also calculated
in a straightfoward manner.
To this end, it is useful to use the following expression of $W^{\dagger}_{x+i}W_x$,
\begin{equation}
W_{x+i}W^{\dagger}_x=\exp [-2\pi qi\sum_y\epsilon_{ij}\nabla_jG(x-y)
(\hat{\rho}(y)-\rho)+i\nabla_i\omega_x].
\label{WW}
\end{equation}
From (\ref{WW}), we get
\begin{equation}
P^{(2)}_W=-V^2_0{\gamma^2\over \beta}\int^{\beta}_0d\tau_1d\tau_2
\langle \exp [-2\pi qi\sum \epsilon_{ij}\nabla_jG(x-y)(\phi^{\dagger}_{y,\sigma}\phi_{y,\sigma}
(\tau_1) -\phi^{\dagger}_{y,\sigma}\phi_{y,\sigma}(\tau_2))]\rangle.
\label{PW}
\end{equation}
The above expectation value $\langle  \exp [\nabla G   \phi^{\dagger}\phi ]   \rangle$ 
is easily evaluated by the hopping expansion.
As $\phi_{x \sigma}$ describes hard-core bosons, the following identity is satisfied for 
an arbitrary c-number $\alpha$,
\begin{equation}
e^{\alpha \phi^{\dagger}\phi}=1+(e^{\alpha}-1)\phi^{\dagger}\phi.
\label{hardcore}
\end{equation}
In the leading order of the hopping expansion, we can use this identity 
to evaluate the following averages;
\begin{eqnarray}
&&\langle \exp[\alpha (\phi^{\dagger}_{x,\sigma}\phi_{x,\sigma}(\tau_1)-
\phi^{\dagger}_{x,\sigma}\phi_{x,\sigma} (\tau_2)) ] \rangle \nonumber\\
&=&\langle [
1+(e^{\alpha}-1)\phi^{\dagger}_{x,\sigma}\phi_{x,\sigma} (\tau_1)]
[1+(e^{-\alpha}-1)\phi^{\dagger}_{x,\sigma}
\phi_{x,\sigma} (\tau_2)]\rangle  \nonumber \\
&=&1. 
\label{expecvalue}
\end{eqnarray}
As a result,  we get
\begin{equation}
P^{(2)}_W=-\gamma^2\beta V_0^2.
\label{P2W}
\end{equation}

From (\ref{Ptree}), (\ref{P2phi}), (\ref{P2b}) and (\ref{P2W}), it is obvious that 
$\chi_0, D_0$, and $V_0$ should develop nonvanishing expectation values 
at sufficiently low $T$, since the coefficients in $P^{(2)}=P_{tree}+
P^{(2)}_{\phi}+P^{(2)}_b+P^{(2)}_W$ become negative.
This result is supported by the higher-order terms which may be calculated
systematically (see  Ref.\cite{IMCSS,IMPFS} for similar calculations).


\subsection{Effective Gauge Theory}

As shown in the previous subsection, the amplitudes of the link fields develop
nonzero expectation values at low $T$.
Therefore one can define the phases of link fields. From the local gauge symmetries (\ref{symmetry1}) and (\ref{symmetry2}), 
the most important degrees of
freedom at low $T$   are these phases.
Actually as we explained before, they can be regarded as gauge fields.
Let us introduce $U(1)$ variables, $U_{xi}$'s, corresponding to them through
\begin{eqnarray}
&&\chi_{x i}=\chi_0U^{ \chi}_{xi}, \;\ D_{x i}=D_0U^D_{xi}, \;\ V_{x i}=V_0U^V_{xi},
\nonumber\\
&& U^{\chi}_{xi}, \; U^D_{xi}, \; U^V_{xi} \; \in U(1). 
\label{gaugefield}
\end{eqnarray}
We substitute (\ref{gaugefield}) into the Hamiltonian of (\ref{H2}),
and integrate out the ``elementary" fields $\phi_{x \sigma}$ and $b_x$ 
by the hopping expansion as in Sect.3.1
in order to obtain an effective Lagrangian of the composite gauge fields, 
$U_{xi}$'s.
This effective Lagrangian must possess the local gauge symmetries corresponding
to $U(1)_{\rm CSS}$ and $U(1)_{\rm PFS}$ of (\ref{symmetry1}) and (\ref{symmetry2}).

The  action $A_{\rm EGT}$ of the effective gauge theory of $U_{xi}$'s
is defined as 
\begin{equation}
\exp (A_{\rm EGT} [U ] )=\int [d\phi][ db][ d\omega] \exp A .
\label{Adef2}
\end{equation}
The hopping expansion below gives (approximately) the following canonical
form with the electric term $A^e$ and the magnetic term $A^m$;
\begin{equation}
A_{\rm EGT}[U ]=A^e+A^m, 
\end{equation}
\begin{eqnarray}
A^e&=&-\int d\tau \sum_{x,i}\Big[\partial_{\tau}U^{\dagger}_{xi}\partial_{\tau}U_{xi}
+\cdot\cdot\cdot
\Big],  \nonumber   \\
A^m&=&\int d\tau \sum_x \Big[U_{x,2}U_{x+2,1}U^{\dagger}_{x+1,2}
U^{\dagger}_{x,1}+\cdot\cdot\cdot  \Big].
\label{action}
\end{eqnarray}
Each hopping term  and   four-point interaction  in $A$ gives rise to
each contribution to the effective action:
\begin{eqnarray}
A^e&=&A^{e}_{\phi}+A^{e}_{b}+A^{e}_{W}+A^{e}_{\phi,4}+A^{e}_{b,4} \nonumber  \\
A^m&=&A^{m}_{\phi}+A^{m}_{b}+A^{m}_{W}+A^{m}_{\phi,4}+A^{m}_{b,4}.  
\label{sumA}
\end{eqnarray}
From the $\phi$-hopping term, we have    
\begin{eqnarray}
A^{e}_{\phi}&=&A^{e}_{\phi,\chi V}+A^{e}_{\phi,D}.
\end{eqnarray} 
The first term is
\begin{eqnarray}
A^{e}_{\phi,\chi V}&=&\Big({3J|\chi_0V_0| \over 8\gamma}\Big)^2\sum_{x,i}\int^{\beta}_0
 d\tau_1 d\tau_2 \bar{U}^{\chi}_{xi}(\tau_1)\bar{U}^{V}_{xi}(\tau_1)U^{\chi}_{xi}(\tau_2)
 U^{V}_{xi}(\tau_2)  \nonumber   \\
 &&\; \times \sum_{\sigma}\langle \phi^{\dagger}_{x,\sigma}(\tau_1)\phi_{x+i,\sigma}(\tau_1)
 \phi^{\dagger}_{x+i,\sigma}(\tau_2)\phi_{x,\sigma}(\tau_2) \rangle \nonumber  \\
 &=&\Big({3J|\chi_0V_0| \over 8\gamma}\Big)^2 \sum_{x,i}{1\over 2}
 (1-\delta^2)\beta^2\sum_n \bar{U}^{\chi}_{xi,n}\bar{U}^{V}_{xi,-n}
 \sum_l U^{\chi}_{xi,l}U^{V}_{xi,-l},
\label{AephiUv}
\end{eqnarray}
where we have introduced Fourier expansions for $U_{xi}(\tau)$'s;
\begin{eqnarray}
&&U_{xi}(\tau)=\sum_ne^{i\omega_n\tau}U_{xi,n},   \nonumber  \\
&&\sum_n U^{\dagger}_{xi,n}U_{xi,n+m}=\delta_{m,0},  
\label{Fourier}
\end{eqnarray}
where $\omega_n=2\pi n/\beta$, $n=0,\pm1,\pm 2,\cdot\cdot\cdot$.
It is obvious from its definition that $U_{xi,n}$ is the amplitude for
those configurations with the  winding number $n$ in the $\tau$-direction.
Therefore, if the static mode $U_{xi,0}$ dominates over all the other oscillating modes
$U_{xi,n\neq 0}$, the gauge dynamics is in the deconfinement phase.
On the other hand, if the amplitudes of  oscillating modes are excitated well
and independently,
then the gauge dynamics is in the confinement phase.
More systematic investigation on this gauge dynamics can be given by the 
Polyakov-Susskind theory \cite{IMCSS,Poly}, 
as we shall see in Sect.4.1.

Similarly, $A^{e}_{\phi,D}$ is evaluated as
\begin{eqnarray}
A^{e}_{\phi,D}&=&{|D_0|^2\over 4}\sum_{x,i}\int^{\beta}_0d\tau_1d\tau_2
\bar{U}^{D}_{xi}(\tau_1)U^{D}_{xi}(\tau_2)  \nonumber  \\
&&\; \times\sum_{\sigma,\sigma'}\langle \tilde{\phi}^{\dagger}_{x+i,\sigma}(\tau_1)
\phi_{x \sigma}(\tau_1)\phi^{\dagger}_{x \sigma'}(\tau_2)
\tilde{\phi}_{x+i,\sigma'}(\tau_2)\rangle \nonumber  \\
&=&{|D_0|^2\over 4}\sum_{x,i}\int^{\beta}_0d\tau_1d\tau_2
\bar{U}^{D}_{xi}(\tau_1)U^D_{xi}(\tau_2)  \nonumber  \\
&&\; \times 2(1+e^{\beta\mu_{\phi}})^{-2}e^{2\mu_{\phi}(\tau_1-\tau_2)}
[\theta(\tau_1-\tau_2)+e^{2\beta\mu_{\phi}}\theta(\tau_2-\tau_1)].
\label{AephiD0}
\end{eqnarray}
The above integral is approximately evaluated as 
\begin{eqnarray}
&&\int^{\beta}_0d\tau_1\int^{\beta}_0d\tau_2e^{2\mu_{\phi}(\tau_1-\tau_2)}
\theta(\tau_1-\tau_2)\bar{U}^{D}_{xi}(\tau_1)U^D_{xi}(\tau_2)  \nonumber  \\
&& \;\; \simeq \int^{\beta}_0d\tau_1\int^{\beta}_0d\epsilon
e^{2\mu_{\phi}\epsilon} \bar{U}^{D}_{xi}(\tau_1)[U^D_{xi}(\tau_1)-\epsilon\partial_{\tau_1}
U^D_{xi}(\tau_1)+{1\over 2}\epsilon^2\partial^2_{\tau_1}U^D_{xi}(\tau_1)]. 
\label{int1}
\end{eqnarray}
The second integral in (\ref{AephiD0}) is evaluated in a similar way;
\begin{eqnarray}
&&\int^{\beta}_0d\tau_1\int^{\beta}_0d\tau_2e^{2\mu_{\phi}(\tau_1-\tau_2+\beta)}
\theta(\tau_2-\tau_1)\bar{U}^{D}_{xi}(\tau_1)U^D_{xi}(\tau_2)  \nonumber  \\
&& \;\; \simeq e^{2\beta\mu_{\phi}}\int^{\beta}_0d\tau_1\int^{\beta}_{0} d\epsilon
e^{-2\mu_{\phi}\epsilon} \bar{U}^{D}_{xi}(\tau_1)\nonumber\\
&&\times [U^D_{xi}(\tau_1)+\epsilon\partial_{\tau_1}
U^D_{xi}(\tau_1) +{1\over 2}\epsilon^2\partial^2_{\tau_1}U^D_{xi}(\tau_1)],
\label{int2}
\end{eqnarray} 
where we have used the periodicity $U^D_{xi}(\tau\pm \beta)=U^D_{xi}(\tau)$.
Collecting (\ref{int1}) and (\ref{int2}), we have
\begin{eqnarray}
A^{e}_{\phi,D}&\simeq&{|D_0|^2\over 2(1+e^{\beta\mu_{\phi}})^2}\sum_{x,i}\int^{\beta}_0d\tau\int^{\beta}_0d\epsilon
\;(e^{2\mu_{\phi}\epsilon}+e^{2\mu_{\phi}(\beta-\epsilon)})  \nonumber  \\
&& \; \times \; [1-{1\over 2}\epsilon^2
\partial_{\tau}\bar{U}^{D}_{xi}(\tau)\partial_{\tau}U^D_{xi}(\tau)].
\label{AephiD}
\end{eqnarray}

For $A^{e}_{b}$,  we have
\begin{eqnarray}
A^{e}_{b}&=&t^2|\chi_0|^2\sum_{x,i}\int^{\beta}_0d\tau_1d\tau_2 \bar{U}^{\chi}_{xi}
(\tau_1)U^{\chi}_{xi}(\tau_2)  \nonumber  \\
&& \; \times\langle b^{\dagger}_{x+i}(\tau_1)b_x(\tau_1)b^{\dagger}_x(\tau_2)
b_{x+i}(\tau_2)\rangle  \nonumber  \\
&=&t^2|\chi_0|^2\delta(1+\delta)\beta^2 \sum_{x,i} \bar{U}^{\chi}_{xi,0}
U^{\chi}_{xi,0}.
\label{Aeb}
\end{eqnarray}

For  $A^{e}_{W}$, the identity (\ref{expecvalue}) is very useful, giving rise to
\begin{eqnarray}
A^{e}_{W}&=&\gamma^2|V_0|^2\sum_{x,i}\int^{\beta}_0d\tau_1d\tau_2
\bar{U}^{V}_{xi}(\tau_1)U^V_{xi}(\tau_2)  \nonumber   \\
&&\; \times \langle W^{\dagger}_x(\tau_1)W_{x+i}(\tau_1)
W^{\dagger}_{x+i}(\tau_2)W_x(\tau_2)\rangle  \nonumber  \\
&=&\gamma^2|V_0|^2\beta^2 \sum_{x,i} \bar{U}^{V}_{xi,0}
U^V_{xi,0}.
\label{AeW}
\end{eqnarray}

It is not straightforward to evaluate the full effect of $A^{e}_{\phi,4}$ and $A^{e}_{b,4}$.
We shall use  perturbative calculation in powers of $\phi^4$ and $b^4$.
To do this, let us rewrite the  $ \phi^4$-term  in the action as
$ \phi^{\dagger}_{x+i,\sigma}\phi_{x+i,\sigma'} 
 \phi^{\dagger}_{x \sigma'}\phi_{x \sigma} =N
[ \phi^{\dagger}_{x+i,\sigma}
\phi_{x+i,\sigma'}] 
N[ \phi^{\dagger}_{x \sigma'}\phi_{x \sigma}] 
+ (\rho/2)(\phi^{\dagger}_{x+i,\sigma}\phi_{x+i,\sigma }
+\phi^{\dagger}_{x \sigma }\phi_{x \sigma})-\rho^2/2$, where
$N[\phi^{\dagger}_{x \sigma'}\phi_{x \sigma}] \equiv \phi^{\dagger}_{x \sigma'}\phi_{x \sigma} -\rho \delta_{\sigma \sigma'} /2$, and similar one  for 
the  $b^4$-term. The second term above is absorbed into the chemical potential term and the
third term is an irrelevant constant.  
Then  we obtain in the leading order,
\begin{eqnarray}
A^{e}_{\phi,4}&=&A^{e}_{4,\chi V}+A^{e}_{4,D},\nonumber\\
A^{e}_{4,\chi V}&=&g^2_{\phi}\Big({3J|\chi_0| \over 8\gamma}\Big)^2|V_0|^2
\sum_{x,i}\int^{\beta}_0
 d\tau_1 d\tau_2d\tau_3 \bar{U}^{\chi}_{xi}(\tau_3)\bar{U}^{V}_{xi}(\tau_3)U^{\chi}_{xi}(\tau_2)
U^V_{xi}(\tau_2)  \nonumber   \\
&&\times\sum_{\sigma} \langle N[\phi^{\dagger}_{x+i,\sigma}\phi_{x+i,\sigma}(\tau_1)]N[\phi^{\dagger}_{x \sigma}
\phi_{x \sigma}(\tau_1)] 
\phi^{\dagger}_{x \sigma}(\tau_3)\phi_{x+i,\sigma}(\tau_3)
 \phi^{\dagger}_{x+i,\sigma}(\tau_2)\phi_{x \sigma}(\tau_2) \rangle \nonumber  \\
 &=&-g^2_{\phi}\Big({3J|\chi_0| \over 8\gamma}\Big)^2|V_0|^2
 \sum_{x,i}{1\over 8}(1-\delta^2)^2\beta^3
\sum_n \bar{U}^{\chi}_{xi,n}\bar{U}^V_{xi,-n}\sum_l U^{\chi}_{xi,l}U^V_{xi,-l}, \label{Ae4Uv} \\
A^{e}_{4,D}&=&g^2_{\phi}{|D_0|^2 \over 4}
\sum_{x,i}\int^{\beta}_0d\tau_1 d\tau_2d\tau_3\bar{U}^D _{xi}(\tau_2)
U^D_{xi}(\tau_3)  \nonumber  \\
&&\times \langle 
N[\phi^{\dagger}_{x+i,\sigma}\phi_{x+i,\sigma'}(\tau_1)]N[\phi^{\dagger}_{x \sigma'}
\phi_{x \sigma}(\tau_1)] \tilde{\phi}^{\dagger}_{x+i,\gamma}(\tau_2)\phi_{x \gamma}(\tau_2)
\phi^{\dagger}_{x \gamma'}(\tau_3)\tilde{\phi}_{x+i,\gamma'}(\tau_3) \rangle  \nonumber  \\
&\sim&-g^2_{\phi}{|D_0|^2 \over 4}\beta^4F(\delta)
\sum_{x,i}\int^{\beta}_0d\tau\;[\partial_{\tau}\bar{U}^D_{xi}(\tau)
\partial_{\tau}U^D_{xi}(\tau)],
\label{Ae4D}
\end{eqnarray}
where $F(\delta)$ is some positive-definite function of  $\delta$
and $g^2_{\phi}={3J \over 8}+\Big({3J\over 8\gamma}|\chi_0|\Big)^2$.
Similarly from the  $b^4$-term, we have
\begin{eqnarray}
A^{e}_{b,4}&=&-{8t^4\over 3J}|\chi_0|^2\sum_{x,i}\int^{\beta}_0d\tau_1d\tau_2d\tau_3
\bar{U}^{\chi}_{xi}(\tau_3)U^{\chi}_{xi}(\tau_2)  \nonumber   \\
&&\times \langle 
N[b^{\dagger}_{x+i}b_{x+i}(\tau_1)]N[b^{\dagger}_{x}b_{x}(\tau_1)] 
b^{\dagger}_{x}(\tau_3)b_{x+i}(\tau_3)
b^{\dagger}_{x+i}(\tau_2)b_{x}(\tau_2) \rangle \nonumber  \\
&=&-{8t^4\over 3J}|\chi_0|^2\delta^2(1+\delta)^2\beta^3\sum_{x,i} \bar{U}^{\chi}_{xi,0}
U^{\chi}_{xi,0}.
\label{Aeb4}
\end{eqnarray}

From (\ref{AephiD}), (\ref{Aeb}) and (\ref{AeW}), it is obvious that,
at high $T$, $\beta \rightarrow 0$, the  coefficients of oscillating modes 
$U_{xi,n\neq 0}$ etc., as well as the static modes, become small, so that
 these modes
are excited randomly, so
both $U(1)_{\rm CSS}$ and $U(1)_{\rm PFS}$ are in disordered-confinement phase.
On the other hand,  at low $T$, $\beta \rightarrow \infty$,
excitations of all the oscillating modes are suppressed,
so both $U(1)_{\rm CSS}$ and $U(1)_{\rm PFS}$ are in the deconfinement phase.
 
We shall see how the other terms would affect the above result.
It is obvious that $A^{e}_{b,4}$ (\ref{Aeb4}) disfavors the 
deconfinement of $U(1)_{\rm CSS}$.
However its effect is small for small $\delta$.
$A^{e}_{\phi,\chi V}$ (\ref{AephiUv}) and $A^{e}_{4,\chi V} $ (\ref{Ae4Uv})
give the couplings between $U^{\chi}_{xi}$ and $U^V_{xi}$.
The sign of the coefficient of this mixing term,
$\sum_n \bar{U}^{\chi}_{xi,n}\bar{U}^V_{xi,-n}\sum_l U^{\chi}_{xi,l}U^V_{xi,-l}$,
depends on the values of $T$, $\chi_0$, etc.
Let us  see the effect of this term.
For example,  let us {\it assume}
that as $T$ goes down, a  phase transition to the deconfinement phase of $U(1)_{\rm CSS}$ takes place  first before  the deconfinement transition of $U(1)_{\rm PFS}$. Then  the static modes $U^{\chi}_{xi,n=0}$ and/or 
 $U^D_{xi,n=0}$ develop nonvanishing
values.
Therefore, one may use a decoupling procedure to evaluate the mixing term as  
\begin{eqnarray}
&&\sum_n \bar{U}^{\chi}_{xi,n}\bar{U}^{V}_{xi,-n}\sum_l U^{\chi}_{xi,l}U^V_{xi,-l}  \nonumber  \\
&\rightarrow& \sum_n \bar{U}^{\chi}_{xi,n}U^{\chi}_{xi,n}\langle \bar{U}^V_{xi,-n}U^V_{xi,-n}\rangle
+\sum_n \bar{U}^{\chi}_{xi,n}U^{\chi}_{xi,n}\langle \bar{U}^{V}_{xi,-n}
U^{V}_{xi,-n}\rangle \nonumber  \\
&\sim&    \langle \bar{U}^{\chi}_{xi,0}U^{\chi}_{xi,0}\rangle
\cdot \bar{U}^V_{xi,0} U^V_{xi,0}.
\label{reducing}
\end{eqnarray}
This implies that, if the coefficient is positive, the deconfinement phase of
$U(1)_{\rm PFS}$ is favored.
On the other hand, if it is negative, the confinement phase is favored.

The megnetic terms of the effective gauge theory are obtained in a 
simlar way.
They play a minor role in the discussion of the 
CD phase transition \cite{IMCSS,Poly}. Their explicit form is 
\begin{eqnarray}
A^{m}_{\phi,\chi V}&=&|\chi_0V_0|^4\Big({3J \over 8\gamma}\Big)^4
\sum_x \int \prod_i d\tau_i  \bar{U}^{\chi}_{x 1}\bar{U}^V_{x 1}(\tau_1)
\bar{U}^{\chi}_{x+1,2}\bar{U}^V_{x+1,2}(\tau_2)  \nonumber  \\
&& \; \times \; U^{\chi}_{x+2,1}U^V_{x+2,1}(\tau_3)U^{\chi}_{x 2}U^V_{x 2}(\tau_4)
\prod^4_{i=1}G_{\phi}(\tau_i-\tau_{i+1})+\mbox{H.c.},
\label{AmphiUv}
\end{eqnarray}
where $\tau_{i+4}=\tau_i$.  \\
Similarly,
\begin{eqnarray}
A^{m}_{b}&=&t^4 |\chi_0|^4
\sum_x \int \prod_i d\tau_i \bar{U}^{\chi}_{x,1}(\tau_1)
\bar{U}^{\chi}_{x+1,2}(\tau_2) U^{\chi}_{x+2,1}(\tau_3)U^{\chi}_{x,2}(\tau_4) 
 \nonumber \\
&& \; \times \;\prod^4_{i=1}G_{b}(\tau_i-\tau_{i+1})+\mbox{H.c.},
\label{Amb}
\end{eqnarray}
\begin{eqnarray}
A^{m}_{W}&=&\gamma^4 |V_0|^4
\sum_x \int \prod_i d\tau_i \bar{U}^{V}_{x,1}(\tau_1)
\bar{U}^{V}_{x+1,2}(\tau_2) U^{V}_{x+2,1}(\tau_3)U^{V}_{x,2}(\tau_4) \nonumber \\
&& \; \times \langle W^{\dagger}_xW_{x+1}W^{\dagger}_{x+1}W_{x+1+2}
W^{\dagger}_{x+1+2}W_{x+2}W^{\dagger}_{x+2}W_x\rangle +\mbox{H.c.} \nonumber  \\
&=&\gamma^4|V_0|^4\sum_x \beta^4\Big(\bar{U}^V_{x,1,0}\bar{U}^V_{x+1,2,0}
U^V_{x+2,1,0}U^V_{x,2,0}\Big)  \nonumber   \\
&& \; \times \; e^{-2\pi iq\rho}[1-\rho+e^{2\pi iq}\rho ]
+\mbox{H.c.}
\label{AmW}
\end{eqnarray}
From (\ref{AmphiUv}), (\ref{Amb}) and (\ref{AmW}), it is obvious that the magnetic 
terms determine the spatial configuration of  $U_{xi}$'s
and the CD picture obtained from the electric terms are usually not affected qualitatively
by the presence of these magnetic terms.

In the following section, we shall give a somewhat detailed study of the phase
structure and phase transitions of the effective gauge theory.

\setcounter{equation}{0}

\section{Phase structure of the effective gauge theory}

In Sect.3.2, we have obtained the explicit form of the effective gauge
theory.
The electric terms $A^{e}_{b}$ and $A^{e}_{W}$ have the following form;
\begin{equation}
\sum_{x,i}\beta^2 \bar{U}_{xi,0}U_{xi,0}.
\label{form}
\end{equation}
From the unitarity condition (\ref{Fourier}), the above term is rewritten as follows;
\begin{eqnarray}
\beta^2 \bar{U}_{xi,0}U_{xi,0}&=&\beta^2\Big(1-\sum_{n\neq 0}\bar{U}_{xi,n}
U_{xi,n}\Big)  \nonumber  \\
&\sim& \beta^2-{2\beta^3\over (2\pi)^2}\int^{\beta}_0 d\tau\partial_{\tau}\bar{U}_{xi}
\partial_{\tau}U_{xi}.
\label{rewrite1}
\end{eqnarray}
The mixing terms $A^{e}_{\phi,\chi V}$ and $A^{e}_{4,\chi V}$ reduce to the form of (\ref{form})
in the vicinity of  CD phase transition or in the deconfinement phase of 
one of the gauge symmetries, because $U^{\chi}_{xi,0}$ 
and/or $U^V_{xi,0}$ dominate over nonstatic modes  (see (\ref{reducing})).
Therefore, essential structure of the CD phase transition is well approximated
by the canonical form of the electric term (\ref{rewrite1}).

In the next subsection, we shall study the CD phase transition of this canonical
gauge theory by mapping it to a classical spin model. The results obtained there
will help us to
follow the detailed study of our effective gauge theory of the t-J model in Sect4.2.


\subsection{Finite-temperature properties of the
canonical gauge theory by a map to the XY model}

Let us consider the   canonical
gauge model, the action  $A(U,V)$ of which is given by
\begin{equation}
A(U,V)= -\int d\tau \sum_{x,i}[a_U\partial_{\tau}\bar{U}_{xi}\partial_{\tau}
U_{xi}+a_V\partial_{\tau}\bar{V}_{xi}\partial_{\tau} V_{xi}],
\label{canonical}
\end{equation}
where we consider two kinds of $U(1)$ gauge fields $U_{xi}$ and $V_{xi}$ and 
{\it single}  gauge symmetry.
For the second field $V_{xi}$,  we use  the same notation   as the link field $V_{xi}$
in Sect.3, but  no confusions should arise since the former  appear only in this 
subsection.  Let $U_{xi}$ and $V_{xi}$ transform under a gauge
transformation under consideration as follows;
\begin{eqnarray}
U_{xi}&\rightarrow& e^{i\alpha_{x+i}}U_{xi}e^{-i\alpha_x},  \nonumber  \\
V_{xi}&\rightarrow& e^{i\alpha_{x+i}}V_{xi}e^{i\alpha_x}.
\label{gaugetrf2}
\end{eqnarray}
It is useful to introduce canonical angles and their  conjugate 
electric operators;
\begin{eqnarray}
&&U_{xi}=\exp (i\theta_{U,xi}), \; \; V_{xi}=\exp (i\theta_{V,xi}),  \nonumber  \\
&&A(U,V)=-\sum_{x,i}[a_U\dot{\theta^2}_{U,xi}+a_V\dot{\theta}^2_{V,xi}],  \nonumber  \\
&&E_{U,xi}=-2a_U\dot{\theta}_{U,xi}\leftrightarrow -i{\partial \over \partial\theta_{U,xi}},  \nonumber  \\  
&&E_{V,xi}=-2a_V\dot{\theta}_{V,xi} \leftrightarrow  -i{\partial \over \partial\theta_{V,xi}}.
\label{electricop}
\end{eqnarray}  
Using the electric field operators (\ref{electricop}), the Hamiltonian $H$ and the 
generator $Q_x $ of the local gauge
transformation (\ref{gaugetrf2}) are given as 
\begin{eqnarray}
H&=&\sum_{x,i}\Big[{1\over 4a_U}E^2_{U,xi}+{1\over 4a_V}E^2_{V,xi}\Big], \nonumber\\
Q_x&=&\sum_i [\nabla_iE_{U,xi}+E_{V,x+i,i}+E_{V,xi}].
\label{gener}
\end{eqnarray}
The partition function $Z$  is the trace over the physical states that are gauge singlet, i.e., satisfy $Q_x =0$. So it is   
written as 
\begin{eqnarray}
Z&=&\mbox{Tr} \prod_x\delta_{Q_x,0}\exp (-\beta H).
\label{partition}
\end{eqnarray}
The trace symbol above implies  the sum over the eigenvalues of
 $E_{U,xi}$ and $E_{V,xi}$ which are integers since $E_{xi}$'s are conjugate to  compact angle variables.
We  introduce Lagrange multiplier angle field $\gamma_x [\in (0, 2\pi)]$ to enforce the gauge constraint $Q_x = 0$. Then $Z$ of
(\ref{partition}) is  given as 
\begin{eqnarray}
Z&=&\prod_{x,i} \sum^{\infty}_{E_{U,xi}=-\infty}\sum^{\infty}_{E_{V,xi}=-\infty}
\int \prod_x {d\gamma_x \over 2\pi}\exp \Big[-\beta H+i\sum_x \gamma_x Q_x\Big] \nonumber\\
&\propto& \int \prod_x {d\gamma_x \over 2\pi}\tilde{\exp} \Big[
-{1\over 2\beta}\sum_{x,i}[{a_U(\nabla_i\gamma_x)^2+a_V(\gamma_{x+i}+\gamma_x)^2}]
\Big],
\label{partition2}
\end{eqnarray}
where we have introduced the periodic Gaussian function,
\begin{eqnarray}
&& \sum^{\infty}_{n=-\infty}\exp (-cn^2+i\gamma n)  \nonumber  \\
&& \hspace{1cm} =\Big({\pi \over c}\Big)^{1/2}\sum^{\infty}_{m=-\infty}
\exp \Big\{-{1\over 4c}(\gamma -2\pi m)^2\Big\}  \nonumber  \\
&& \hspace{1cm} \equiv \Big({\pi \over c}\Big)^{1/2}\tilde{\exp} \Big(-{1\over 4c}
\gamma^2\Big).
\label{pGauss}
\end{eqnarray}

The last expression of (\ref{partition2}) is nothing but the partition function
of a classical XY spin model with a ``double"-Villain action.
This spin model has a global $Z_2$ symmetry under $\gamma_x \rightarrow 
\gamma_x + \pi$, and
belongs to the same universality class as the
following XY model with ``double"-cosine action, or ``anisotropic'' XY model,
\begin{eqnarray}
&&Z_{\rm XY}=\int \prod_x{d\gamma_x\over 2\pi}\exp \Big[J_1\sum_{x,i}
\cos (\gamma_{x+i}-\gamma_x)+J_2\sum_{x,i}\cos (\gamma_{x+i}+\gamma_x)\Big]
\nonumber  \\
&&\hspace{0.85cm} =\int \prod_x{d\gamma_x\over 2\pi}\exp \sum_{x,i}\Big[
(J_1+J_2)\cos  \gamma_x  \cos  \gamma_{x+i} +(J_1-J_2)\sin  \gamma_x 
\sin  \gamma_{x+i} \Big],  \nonumber  \\
&& \; J_1\equiv \beta^{-1}a_U, \; \; J_2\equiv \beta^{-1}a_V.
\label{XYmodel}
\end{eqnarray}
By the MF theory \cite{IMCSS} and also by the numerical calculations \cite{SM}, 
it is shown that
the model has  three phases;
\begin{enumerate}
\item Disordered phase for $J_1+J_2 \ll 1$,   \\
The spin-spin correlation functions behave  as 
\begin{eqnarray}
\langle e^{i\gamma_x}e^{-i\gamma_0}\rangle &\sim& e^{-c|x|}, \label{spinspin1}  \\
|x| &\rightarrow& \infty
\nonumber  
\end{eqnarray}
\item Quasi-long-range-ordered phase for $J_1 \gg 1, \; J_2=0$ or 
$J_1=0, \; J_2 \gg 1$,   \\
This is the well-known Kosterlitz-Thouless (KT) phase of the $O(2)$-invariant
classical XY spin model in two-dimensions; 
\begin{eqnarray}
\langle e^{i\gamma_x}e^{-i\gamma_0}\rangle &\sim& |x|^{-c}, \label{spinspin2}  \\
|x| &\rightarrow& \infty
\nonumber  
\end{eqnarray}
or
\begin{eqnarray}
\langle e^{(-)^xi\gamma_x}e^{-i\gamma_0}\rangle &\sim& |x|^{-c}, \label{spinspin2'}  \\
|x| &\rightarrow& \infty
\nonumber  
\end{eqnarray}
\item Ordered phase for $J_1+J_2 \gg 1, J_1\neq 0,\; J_2\neq 0$, \\
The $Z_2$ symmetry is spontaneously broken in this phase, and the ground state
is the configuration with $\gamma_x=0$ (mod $\pi$).
\begin{eqnarray}
\langle e^{i\gamma_x}e^{-i\gamma_0}\rangle &\sim& \mbox{const.}+e^{-c|x|}. \label{spinspin3}  \\
|x| &\rightarrow& \infty
\nonumber  
\end{eqnarray}
\end{enumerate}

Let us put a pair of oppositely-charged static sources at $x=0$ and $x=R$
in the original gauge theory.
In this case, the constraint changes as $Q_x=\delta_{x,0}-\delta_{x,R}$.
It is easily seen \cite{Poly} that the potential energy $W(R)$ of this state
  is related with
the above spin-spin correlation functions as follows;
\begin{equation}
\langle e^{i\gamma_R}e^{-i\gamma_0} \rangle =\exp [-\beta W(R)].
\label{WR}
\end{equation}
Therefore, 
\begin{enumerate}
\item Disordered phase: $W(R) \propto R$. \\
This phase corresponds to the confinement phase.
quasiexcitations are only charge-neutral compounds.
\item Quasi-long-range ordered phase: $W(R) \propto \ln R$.  \\
This phase corresponds to the massless Coulomb phase.
Because of the long-range interaction by the massless gauge boson,
some nontrivial infrared behaviors are expected.
Resummation method of the perturbative expansion or renormalization-group 
study are useful to  investigate this phase \cite{RG}.
\item Ordered phase: $W(R) \propto e^{-cR}$.  \\
This phase corresponds to the Higgs phase with a  massive gauge boson.
There is only short-range interaction between charged particles.
\end{enumerate}
The above result is qualitatively understood from the fact that $\gamma_x$
is the Lagrange multiplier for the constraint of the local-gauge invariance.
In the disordered phase of the spin system, $\gamma_x$ fluctuates strongly.
This means that the local constraint in the original gauge theory
operates quite strictly.
On the other hand, in the ordered phase of the spin system, $\gamma_x$ has a
nontrivial expectation value and fluctuations around it are very small.
In the original gauge theory, this mean that the local-gauge invariance
is not faithfully respected by quasiexcitations.
Charged particles can appear as the physical excitations.
The phase of the quasi-long-range order is in between.

One might conceive the possibility of mixed  dynamics; i.e.,
 the $U$ field (say, for the $b-\bar{b}$ channel) is in confinement phase
whereas the $V$ ( for the $b-b$ channel) field is in deconfinement (or the other way
around).
To study the dynamics of $V$ field, one needs to calculate the potential energy
of a pair of sources of {\it same charges}, hence the 
spin correlation functions, 
$\langle\exp(i(-1)^{x_1 +x_2} \alpha_x) \exp(i\alpha_0)\rangle$.
By  using a MF theory and the high-temperature expansion for (\ref{XYmodel}), 
it is straightforward to see that  two sets of correlation functions,
$\langle\exp(i\alpha_x) \exp(i\alpha_0)\rangle$ for the $U$-dynamics and
$\langle\exp(i(-1)^{x_1 +x_2} \alpha_x) \exp(i\alpha_0)\rangle$ for
the $V$-dynamics, 
behave similarly at large $|x|$. 
Therefore two gauge dynamics are
always in the same phase and such mixed phases do not exist.


\subsection{CSS and PFS  in the double-boson t-J model }

In the previous subsection, we examined the phase structure of the canonical
gauge theory.
It is rather straightforward to apply these results to the 
effective gauge theory of the t-J mode
in the double-boson representation, which we derived in Sect.3.
By using (\ref{rewrite1}), $A^{e}_{b}$ of (\ref{Aeb}) and  
$A^{e}_{W}$ of (\ref{AeW}) can be rewritten as follows;
\begin{eqnarray}
A^{e}_{b}&\simeq&-t^2|\chi_0|^2{2\over (2\pi)^2}\delta(1+\delta)\beta^3
\int d\tau \sum_{x,i}
\partial_{\tau}\bar{U}^{\chi}_{xi}\partial_{\tau} U^{\chi}_{xi}, \nonumber  \\
A^{e}_{W}&\simeq&-\gamma^2|V_0|^2{2 \over (2\pi)^2}\beta^3  \int d\tau \sum_{x,i}
\partial_{\tau}\bar{U}^V_{xi}\partial_{\tau} U^V_{xi}.
\label{effect1}
\end{eqnarray}
Similarly, the kinetic term (\ref{AephiD}) of $U^D$ becomes 
\begin{equation}
A^{e}_{\phi,D}\simeq -|D_0|^2g(\delta)\beta^3\int d\tau \sum_{x,i}
\partial_{\tau}\bar{U}^D_{xi}\partial_{\tau}U^D_{xi},
\label{effect2}
\end{equation}
where
\begin{eqnarray}
g(\delta)&=&{1 \over 4(1+e^{\beta\mu_{\phi}})^2\beta^3}\int^{\beta}_0
d\epsilon \; (e^{2\mu_{\phi}\epsilon}+e^{2\mu_{\phi}(\beta-\epsilon)})\epsilon^2,
\end{eqnarray}
and a similar expression for $A^{e}_{4,D}$ of (\ref{Ae4D}).

Let us assume that, as lowering $T$,   the  CD phase transition of 
$U(1)_{\rm CSS}$ 
occurs first before  the CD transition of $U(1)_{\rm PFS}$, i.e., $T_{\rm CSS}
> T_{\rm PFS}$.
Then we  determine the transition temperature $T_{\rm CSS}$ as follows.
In this case, the mixing terms (\ref{AephiUv}) and (\ref{Ae4Uv}) give 
only minor effects, because $U(1)_{\rm PFS}$ 
is still in a disordered-confinement phase 
at $T = T_{\rm CSS}$, and the associated gauge field $U^V_{xi}$ 
fluctuates rather randomly. Explicitly, the decoupled term  (\ref{reducing})
is negligible since $\langle\bar{U}^V_{xi,0}U^V_{xi,0}\rangle$ is small.
Therefore, from (\ref{effect1}) and  (\ref{effect2}), we estimate  
\begin{eqnarray}
a_U&=&2t^2|\chi_0|^2{\delta(1+\delta)\over (2\pi)^2}\beta^3, \nonumber\\
a_V&=&|D_0|^2g(\delta)\beta^3,
\nonumber 
\end{eqnarray}
in (\ref{canonical}).
From the mapping (\ref{XYmodel}) considered in Sect.4.1, the couplings $J_1$ and $J_2$
in the mapped XY spin model are given by
\begin{eqnarray}
J_1&=&2t^2|\chi_0|^2{\delta(1+\delta)\over (2\pi)^2}\beta^2,\nonumber\\
J_2&=&|D_0|^2g(\delta)\beta^2.
\label{effectJ}
\end{eqnarray}
The phase transition line is approximately given by $J_1+J_2=1$.   
 At high $T$ we have $J_1 + J_2 <1$ ,  since the magnitudes $\chi_0,D_0,V_0$ are 
certainly decreasing functions of $T$, so
 the $U(1)_{\rm CSS}$ is in the confinement phase.
At low $T$, $J_1 + J_2 > 1$ and it is in the deconfinement phase, as we assumed.
More precisely, there are two kinds of deconfinement phase.
For $\chi_0\neq 0$ and $D_0\neq 0$, the Higgs phase appears.
If one of $\chi_0 $ and $D_0 $ vanishes, then the Coulomb phase appears.
As we explained in Sect.3.2, this latter transition ($D_0 =0$) 
corresponds to the CSS
(see Ref.\cite{IMCSS} for more detailed discussion), and   $T_{\rm CSS}$
 is estimated as \cite{cfsb}
\begin{equation}
T_{\rm CSS}\simeq \Bigg( 2t^2|\chi_0|^2{\delta(1+\delta) \over (2\pi)^2}+|D_0|^2
g(\delta)\Bigg)^{1/2}. 
\label{TCSS}
\end{equation}

At  $ T< T_{\rm CSS}$, the static component of $U^{\chi}_{xi}$
dominates, $U^{\chi}_{xi}(\tau)\sim U^{\chi}_{xi,0}$.
From (\ref{reducing}), the mixing terms $A^{e}_{\phi, \chi V}$ 
and $A^{e}_{4,\chi V}$ 
give similar  effects upon the dynamics of $U^V_{xi}$ as for the previous case
 (\ref{effect1}).
In order to estimate them, we have to know the quantitative behavior of 
$\chi_0$, $V_0$, etc.
Such a quantitative investigation of them is under study and will appear elsewhere. 
It is however  
 expected that, at some parameter region of the $ T-\delta $ plane, 
the deconfinement phase of $U(1)_{\rm PFS}$ is realized.

On the contrary, if we assume that the CD transition of $U(1)_{\rm PFS}$ takes place 
at higher $T$ than 
that of $U(1)_{\rm CSS}$, one obtains similar results, 
but the roles of  $U(1)_{\rm CSS}$ and $U(1)_{\rm PFS}$ are interchanged.

Let us consider the phase structure {\em near the half filling}.
Some qualitative but still interesting  results are obtained there without 
knowledge of detailed behavior of the amplitudes. Let us consider the region
$T < T_{\rm CSS}$.
Near the half filling $\delta \sim 0$, the coefficient of $A^{e}_{b}$ of (\ref{Aeb}) is very small and so
the zero mode $U^{\chi}_{xi,0}$ 
has a finite but {\em very small} expectation value $\langle \bar{U}^{\chi}_{xi,0}U^{\chi}_{xi,0}\rangle$ even if 
the gauge dynamics of $U(1)_{\rm CSS}$ is in the deconfinement phase as we assumed.
This leads   that the mixing 
terms $A^{e}_{\phi,\chi V}$ and $A^{e}_{4,\chi V}$
play no important role as the decoupling (\ref{reducing}) shows.  
On the other hand, the term $A^{e}_{\phi,D}$ of (\ref{effect2}) 
has a finite coefficient as  
$\delta \rightarrow 0 (\mu_{\phi} \rightarrow 0 )$,
$$
g(\delta=0)={1 \over 48}.
$$
The coefficient of the term $A^{e}_{4,D}$ of (\ref{Ae4D}) has a similar behavior.
Thus, for $D_0\neq 0$, the zero modes $D_{xi,0}$ develop at low $T$. 
$T_{\rm CSS}$ is estimated by (\ref{TCSS}).
Concerning to the PFS, we apply the result of Sect.4.1 to the 
$U^V_{xi}$-dynamics of the term  $A^{e}_{W}$ (\ref{effect1}).
The condition $J_1 +J_2 =1 (J_2 =0)$ reads 
$2\gamma^2|V_0|^2(2\pi)^{-2}\beta^2 =1$, so it predicts the
CD transition at 
\begin{equation}
T_{\rm PFS} \simeq \frac{\gamma}{\sqrt{2}\pi} |V_0| .
\label{TPFS}
\end{equation}
The PFS takes place at  $T < T_{\rm PFS}$. 
Therefore, at sufficiently low $T$, 
the effective gauge theory should be in the phase of
($U(1)_{\rm CSS}$, $U(1)_{\rm PFS}$)=(deconfinement, deconfinement), 
whose physical picture is discussed in Sect.2.2.
In particular, the spinons behave as bosons.
A simple MF study is applicable, and
existence of a long-range antiferromagnetic order is expected 
at low $T$ (at $T=0$ in the exactly two-dimensional case).
Here we point out that, if the condition $2 (\gamma |V_0| /2\pi)^2  > |D_0|^2g(0)$ 
is satisfied,   the expressions (\ref{TCSS}) and (\ref{TPFS})  
 show that  the phase like ($U(1)_{\rm CSS}$, $U(1)_{\rm PFS}$)=(confinement,
deconfinement) is realized at  $ T_{\rm CSS} < T < T_{\rm PFS}$.
As explained in Sect.2.2, if  holons cannot move by some localization effect,
this phase  corresponds to an insulator with
{\em massive} spin excitations.

Now let us consider the other case of  intermediate hole concentrations.
 Here $A^{e}_{b}$ is dominant due to the large coefficient 
 and the associated gauge field $U^{\chi}_{xi}$ becomes static at low $T$.
In this case, besides $A^{e}_{b}$ and $A^{e}_{W}$, the $\phi^4$-term and 
 $A^{e}_{4,\chi V}$ play  an important role.
From (\ref{AephiUv}) and (\ref{Ae4Uv}), we see that the term $A^{e}_{4,\chi V}$
dominates over $A^{e}_{\phi,\chi V}$ at low $T$,  $T < {1\over 4}g^2_{\phi}(1-\delta^2)$.
As one can see from (\ref{reducing}) and the fact that the coefficient
in $A^{e}_{4,\chi V}$ of (\ref{Ae4Uv}) is negative, the  term $A^{e}_{4,\chi V}$  {\it disfavors} co-existence of   two sets of the 
static modes, $U^{\chi}_{xi,0}$ and $U^V_{xi,0}$.
This indicates that if $T_{\rm CSS} > T_{\rm PFS}$, the CSS takes place at low $T$, but    the PFS is suppressed by the $\phi^4$-interaction.
Therefore, the phase ($U(1)_{\rm CSS}$, $U(1)_{\rm PFS}$)=
(deconfinement, confinement) is expected. This concludes that
 the spinons are bosonic at intermediate $\delta$'s. 
The importance of  4-point interactions in separation phenomena  was recognized in the PFS of FQHE.
Actually, we remember that, for quasiexcitations in the half-filled Landau level, the Coulomb repulsion
is essential for the stability of  composite fermions \cite{IMPFS}.

Let us discuss the effect of magnetic terms at intermediate $\delta$'s 
in some detail.
It is well known that {\it fermions} under a constant magnetic field $B$
lower their energy drastically when $B$ is just a unit flux per fermion, since
each fermion may absorb a flux quantum to convert itself to a single boson and 
these bosons may condense in the  lowest energy state.
Statistical transmutation of electrons in FQHE at $\nu = 1/(2n+1)$ is an example.
However, in our double-boson representation, we face {\it bosons} in  
a magnetic  field and no  statistical transmutation occurs in contrast with the case of
fermions. Instead we can argue the relative stability of CSS and PFS.
The magnetic term $A^m_b$ (\ref{Amb}) favors the uniform fluxless configuration
of $U^{\chi}_{xi}$, whereas $A^m_W$ (\ref{AmW}) favors 
spatial-flux configuration
of $U^V_{xi}$ with $2q$ flux quanta for spinon.
From this difference in fluxes which a $b$-particle and a $\phi$-particle
feel, one may argue that
the CSS is more stable than the PFS, i.e., $T_{\rm CSS} > T_{\rm PFS}$.
For {\it bosons} in a constant 
magnetic field $B$, there appear the same one-particle
Landau levels as in the case of fermions. 
However, all the bosons can occupy the lowest state, 
the energy of which is the zero-point oscillation $\propto B^{1/2}$.
Let us assume that only the CSS takes place. Then neither of
the quasiexcitations, the $b$-particles nor the composites 
$f_{x\sigma}=W_x^{\dagger} \phi_x$ 
feel any finite magnetic field, since  the CS fluxes, $W_x^{\dagger}$,  feel the 
field $\bar{U}^V_{xi}$ which cancels $U^V_{xi}$ for $\phi_x$'s.
There are no zero-point energy. On the contrary, let us assume
only the PFS takes place. Then each quasiexcitation, $b^{\dagger}_x  \phi_{x\sigma}$, feels
$2q$ flux and so stores zero-point energy.
This leads us to the above conclusion of relative stability.
 
Anyway, as we explained in Sect.2.2, the region in question 
corresponds to the anormalous metallic phase
of high-Tc cuprates.
On the other hand, at sufficiently high $T$, it is obvious that both $U(1)_{\rm CSS}$
and  $U(1)_{\rm PFS}$ are in the confinement phase, 
and the quasiexcitations are the original electrons.
This result is also in agreement with experiments.

In Fig.1 we present a possible phase diagram in the $T-\delta$ plane.
The line of $T_{\rm CSS}$ is based on the numerical calculations in Ref.\cite{IMS}.
The line of $T_{\rm PFS}$ is drawn assuming that it is a 
decreasing function of $\delta$.  
This is expected from $A^e_{4,\chi V}$ (\ref{Ae4Uv}) and the 
fact that the amplitude $\chi_0$ increases very rapidly as a function of $\delta$ \cite{IMS}.
Also in most of the interesting region of $\delta$, the estimated $T_{\rm CSS}$ is
smaller than $J$ or $g^2_{\phi}$.
Therefore the region below $T_{\rm CSS}$
is partitioned into two region, the low-$\delta$ region named the region (I), and the 
high-$\delta$ region, the region (II). 
These regions correspond to
($U(1)_{\rm CSS}$, $U(1)_{\rm PFS}$)=(deconfienement, deconfinement),
and  (deconfienement, confinement), respectively.
The spinons are bosonic in the former region,
while they are fermionic in the latter region.
As explained above, in the region (I) near the half filling, the terms in the 
effective gauge theory, $A^e_{\phi,D}$, $A^e_{4,D}$ and $A^e_W$, are dominant.
In the region (II) of the intermediate hole concentration, the terms
$A^e_b$, $A^e_W$ and $A^e_{4,\chi V}$ are effective.

\setcounter{equation}{0}
\section{Conclusions}

In this paper, we have studied nature of the quasiexcitations of the two-dimensional 
t-J model, starting from the slave-boson representation.
Especially we are interested in   when the
 statistics of quasiexcitations transmute as $T$ and $\delta$ change
and how they are related to the CSS.
To this end, we used a general gauge-theoretical approach to 
separation phenomena   developed in the previous papers \cite{IMCSS,IMPFS} for 
two independent separation phenomena, CSS and PFS.

The CSS is understood as a CD phase transition of the gauge fields,  
the phase degrees of freedom of the link ``mean fields" expressing the correlations
among nearest-neighbor holons  and spinons.
In Ref.\cite{IMCSS,IMS},   the 
critical temperature $T_{\rm CSS}$ of the CD transition is calculated. 

To study statistics of quasiexcitations, the CS gauge theory is a suitable apparatus.
This problem is closely related to the idea of  CF in the half-filled
Landau level.
In that case, however, an electron  transmutes to another fermion,  a CF,
by  attaching two (or even number of) CS flux quanta.
In the present t-J model a fermionic spinon  in the slave-boson formalism is
viewed as a composite of  a bosonic spinon and odd number of CS flux quanta.
The possibility of dissociation of a spinon into  a bosonic spinon and CS flux, i.e., PFS, implies the change
of statistics of spinons.   
As in the case of CSS, we  calculated
the transition temperature $T_{\rm PFS}$.
 
Explicitly, we first categorized the possible phases of the t-J model based on  the
two  local gauge symmetries of the system, $U(1)_{\rm CSS}$ and $U(1)_{\rm PFS}$.
By using the hopping expansion, we obtained the explicit form of the
effective gauge theory.
Then, by mapping this effective theory to the classical (an)isotoropic
XY spin model in two dimensions and using the known results on this model,
we identified the physical nature of each phases.
The result is summerized in Fig.1.
Especially, our investigation indicates that, near the half filling and at low $T$, 
 the PFS takes place and spinons behave  like bosons rather than fermions.

For more quantitative study, detailed MF calculations are required.
In the framework of  slave-boson formalism, this is carried out  and  
a line of the CSS transition in the $T-\delta$ plane is obtained \cite{IMS}.  
What is interesting is that $T_{\rm CSS}$ is only $10\sim 20$ \% of the MF 
critical temperature. This exhibits that the effects of gauge-field fluctuations
are very important not only quantitatively but also qualitatively.  

It is also interesting and complementary to the present study to start with
the slave-fermion representation and then move to the double-boson
(or double-fermion)
formalism using the CS gauge theory.
Parallel discussion is possible in this formalism and we expect nontrivial and 
interesting result which will be reported elsewhere. 
In particular it will shed some light on the statistics  of holons; a 
topic that we do not discuss sufficiently in the present paper. 
One may conceive
yet other representations to start with, although they might look excentric,
including  anyonic excitations.
We plan to pursue the problem of statistics in the t-J model by using techniques 
based on gauge theory similar to the present ones, but from a more 
general point of view in scope of  these different starting representations.

Concerning to the  statistics of  quasiexcitations in the t-J model, 
there are  some recent works \cite{others}.  However, the present approach is
distinguished from them by applying the knowledge of strong-coupling 
gauge theory. We stress that the notion of confinement and deconfinement
is powerful to understand  separation phenomena of degrees of freedom in
condensed-matter physics in an intuitive and coherent manner. 

In this paper, we
showed explicitly how the gauge-theoretical approach is used to
the {\it combined situation} of two typical separation phenomena, CSS and PFS.
According to the gauge theory of CD transitions, we expect  two genuine
phase transitions of KT type;  the CSS transition   at $T_{\rm CSS}$ where
the CSS takes place, and the PFS transition at $T_{\rm PFS}$  where the statistics of
spinons changes.  These should be taken seriously as  well as the 
experimentally established
transitions; the antiferromagnetic transition at $T_{\rm Neel}$ \cite{YTIM} and
the superconducting transition at $T_c$ \cite{imsuper}.
Sufficient experimental indications seem to  
have appeared for the CSS
\cite{IMS}.  It seems  remaining  to find such indications for PFS.

\eject

\eject
FIGURE CAPTION

\bigskip
FIG.1. A possible phase diagram in the $T-\delta$ plane.
The region below $T_{\rm CSS}$
is partitioned into two region, (I) the low-$\delta$ region where the
 PFS occurs, and  
(II) the high-$\delta$ region without PFS.
The spinons are bosonic in (I), and fermionic in (II).
In Ref.\cite{IMS}, $T_{\rm CSS}$ is explicitly evaluated as a function of $\delta$ by   numerical calculations. These two curves $T_{\rm CSS}$
and $T_{\rm PFS}$  should be 
taken seriously as well as the experimentally established curves;
the antiferromagnetic transition line $T_{\rm Neel}$ \cite{YTIM} and the
superconducting transition line $T_c$ \cite{imsuper}.


\begin{thebibliography}{99}
\bibitem{decoupling}
In the MF theory \cite{meanfield1}, Ubbens and Lee
ignored the  $-\frac{1}{4}n_x n_{x+i}$ term in
the starting t-J Hamiltonian (\ref{HtJ}).
Their decoupling procedure gives rise to the coefficient
$\frac{2}{3J}$ in front of $|D_{xi}|^2$ instead of the present one,
$\frac{1}{2J}$. Our choice of the coefficient is assured
in path-integral formalism  to reproduce (\ref{Hsb}).
Their MF Hamiltonian  ignored the counter terms,
the $b^4$-term and $f^4$-term in (\ref{Hdec1}),
which are necessary to maintain the equivalence to (\ref{Hsb}).
\bibitem{note1}
The present Hamiltonian includes the "counter terms", the $\phi^4$-term and   
the $b^4$-term.   
They are necessary to maintain the equivalence to the 
original t-J Hamiltonian since they work as the counter terms to cancel
unwelcome terms produced by the Hubbard-Storatonovich transformation. 
In the MF analysis of Ref.\cite{meanfield1}, they are ignored, but
they play an important role in the CSS and PFS as shown in Sect.4.2.
\bibitem{sbsftrans}It is easily verified that spinon and holon operators 
in the slave-boson and slave-fermion formalisms are related with each other as follows;
$\phi_{x \sigma}=e^{-iq \int d^2y\theta (x-y)
\hat{\rho}_{tot}(y)}f_{x \sigma}, \;
\psi_x=e^{-iq \int d^2y\theta (x-y)\hat{\rho}_{tot}(y)}b_x$, where $\psi_x$ is the holon operator
in the slave-fermion representation and $\hat{\rho}_{tot}(x)
=\sum_{\sigma}f^{\dagger}_{x \sigma}
f_{x \sigma}+b^{\dagger}_xb_x=\sum_{\sigma}\phi^{\dagger}_{x \sigma}\phi_{x \sigma}
+\psi^{\dagger}_x\psi_x$.
\bibitem{ps}In the present paper, we are not considering the possiblity of the phase
separation.
\end{thebibliography}
\end{document}